\documentclass[11pt,twoside]{book} 
\usepackage{asp2008n}
\usepackage{times}
\usepackage{lscape}
\usepackage{epsf}

\usepackage{graphicx}
\usepackage{amssymb}
\usepackage{longtable}
\usepackage[figuresright]{rotating}
\usepackage{multirow}

\newcommand{\simless}{\mathbin{\lower 3pt\hbox {$\rlap{\raise 5pt\hbox{$\char'074$}}\mathchar"7218$}}}

\mainmatter

\newlength{\deftabcolsep}
\begin{document}

\bibliographystyle{aa}
\thispagestyle{myheadings}

\title{Young Stars and Clouds in Camelopardalis}

\author{V. Strai\v{z}ys and V. Laugalys}
\affil{
Institute of Theoretical Physics and Astronomy, Vilnius University,
  Go\v stauto 12, Vilnius LT-01108, Lithuania}

\begin{abstract} Star formation in the Local spiral arm in the direction
of the Galactic longitudes 132--158$\deg$ is reviewed.  Recent
star-forming activity in this Milky Way direction is evidenced by the
presence here of the Cam OB1 association and dense dust and molecular
clouds containing H$\alpha$ emission stars, young irregular variables
and infrared stellar objects.  The clouds of the Local arm concentrate
in two layers at 150--300 pc and $\sim$\,900 pc from the Sun. The
Perseus arm objects in this direction are at a distance of about 2 kpc.
\end{abstract}

\section{Introduction}   

At the border of Cassiopeia and Camelopardalis the Milky Way loses its
brightness and almost disappears at the Galactic longitude $\ell$ =
143$\deg$.  Its brightness comes back only at $\ell$ = 170$\deg$, near
the Auriga border.  This Milky Way discontinuity most naturally can be
explained as blocking of light of distant stars by interstellar dust
clouds.  The presence of dense dust and molecular clouds in the Local
arm in the direction of Camelopardalis now is confirmed both by optical
and radio observations which will be reported in the subsequent
sections.  We will also discuss star formation in Camelopardalis which is
evidenced by the presence here of the Cam OB1 association at a distance
of 0.9--1.0 kpc and by other young objects.

The population of dust and molecular clouds, as well as of young stars,
will be analyzed in the ranges of Galactic longitudes 132--158$\deg$ and
latitudes $\pm$\,12$\deg$.  These ranges are defined by the distribution
of the supposed Cam OB1 association members.  The region is located
mainly in Camelopardalis, but also includes edges of Cassiopeia, Perseus
and Auriga.  Since our attention will be concentrated only on the
objects related to star formation in the Local spiral arm, distant objects
in the Perseus arm will be touched on only occasionally.  Objects in the
nearby areas of Cassiopeia are reviewed by  Megeath et al. and by
Kun in this book.

\section{Nebulae and Clusters of the Local Arm} 

\begin{figure}[!th]
\centering
\includegraphics[draft=False]{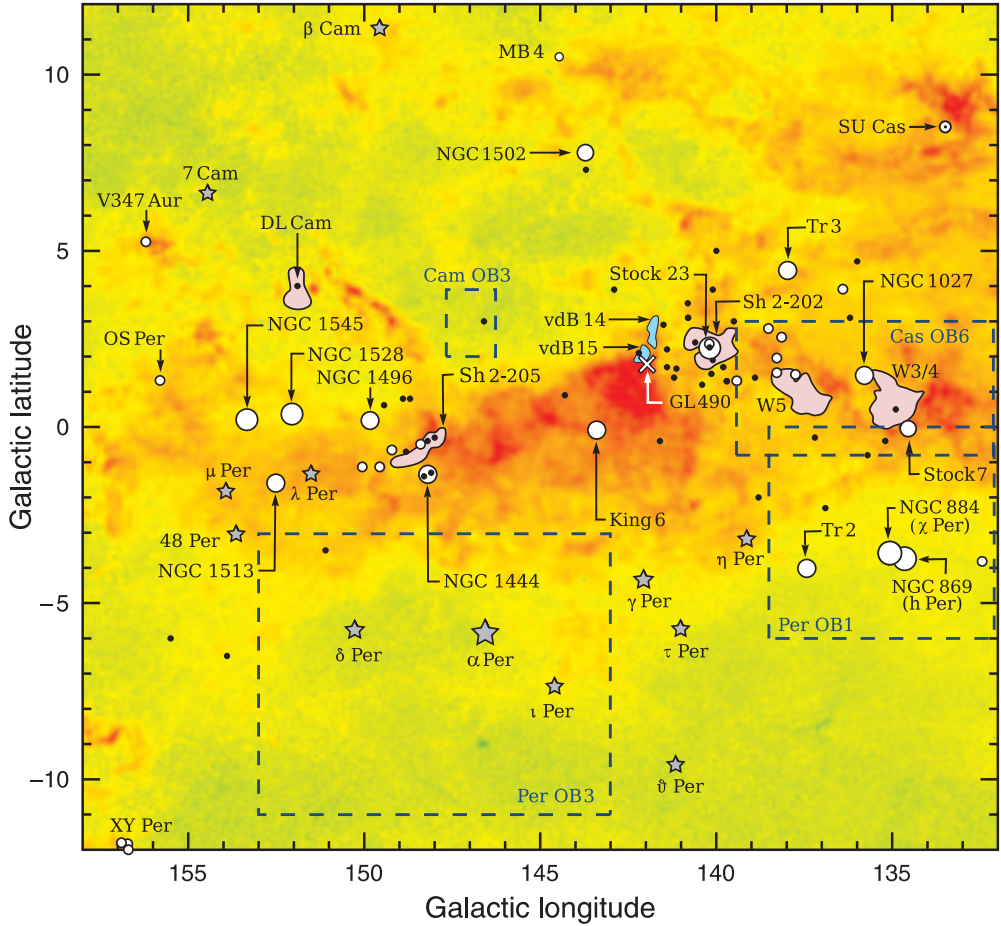}
\vspace{0mm}
\caption{Dust clouds in Camelopardalis from Dobashi et al.  (2005)
plotted together with the clusters and nebulae and the known young
objects.  White large circles designate positions of 12 open clusters
belonging to the Local arm and of the double cluster h+$\chi$ Per.  The
bright rosy patches designate the emission nebulae W3/4, W5, Sh\,2-202,
Sh\,2-205 and DL Cam, the two small blue patches are the reflection
nebulae vdB\,14 and vdB\,15.  Black dots designate the association Cam
OB1 members of spectral types O--B3 V--III and supergiants, small white
circles designate young
irregular variables and H$\alpha$ emission stars.  The YSO GL\,490 is
shown as a white cross.  Dashed lines delimit the areas of the
associations Cas OB6, Per OB1, Cam OB3 and Per OB3. Star-like symbols
designate the Camelopardalis and Perseus stars brighter than $V$ = 4
mag.}
\end{figure}

Figure 1 gives a map of the reported area in Galactic coordinates
indicating some of the objects mentioned in this review.  In comparison
to Cassiopeia, the Camelopardalis section of the Milky Way is relatively
poor in bright nebulae:  here only a few faint emission and reflection
nebulae are present.  At $\ell$,\,$b$ = 140$\deg$, +2.0$\deg$ the
emission nebula Sh\,2-202 of $\sim$\,2$\deg$ diameter is seen (Sharpless
1959).  Located at a distance of 800 pc (Fich \& Blitz 1984), the nebula
is ionized by the nearby star HD 19820 (O8.5\,III).  A group of bright
stars inside it is known as the open cluster Stock 23 with a blue nebula
around (Figure 2).  According to Kharchenko et al.  (2005), this cluster
is not related to the nebula and is a foreground object at a distance of
380 pc.  However, in the $V$ vs.  $B$--$V$ diagram stars of this group
are so scattered that the cluster seems to be not real.  One of these
stars, HD 20134 (B2\,IV), is in the list of the Cam OB1 association
members.

The two elongated reflection nebulae, vdB\,14 and vdB\,15, each
extending for 0.7--0.8$\deg$ in the N--S direction, are located at
$\ell$, $b$ = 142$\deg$, 1.8--3.0$\deg$ (Figure 3).  They are
illuminated by two supergiants, HD 21291 (B9\,Ia) and HD 21389 (A0\,Ia),
respectively.  Racine (1968) suggested that these two stars, and three
other young stars in the vicinity, form a separate association Cam R1.
However, there is no ground to isolate these stars from other association
Cam OB1 members.

\begin{figure}[tb]
\centering
\includegraphics[width=0.85\textwidth,draft=False]{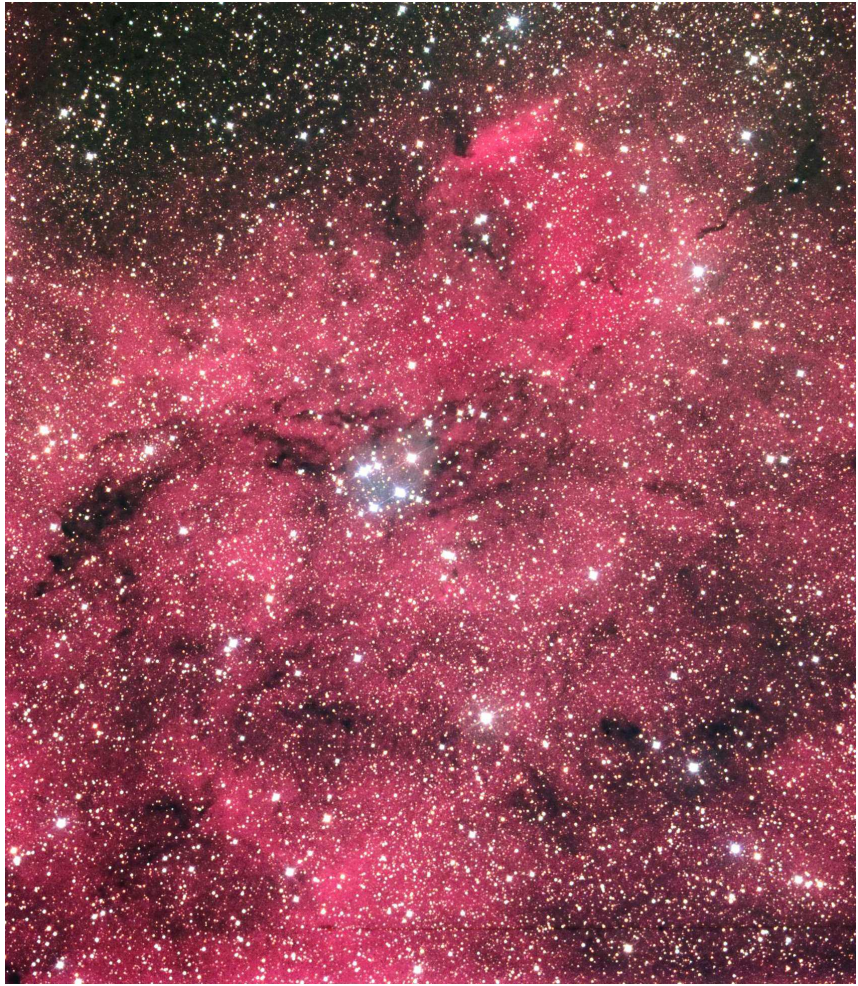}
\vspace{0mm}
\caption{The emission nebula Sh\,2-202 and the doubtful cluster
Stock 23 in the center. North is up
and east is left. The figure is $\sim$\,2.2$\deg$ high. Courtesy
Dean Salman.}
\end{figure}

At $\ell$, $b$ = 148--149$\deg$, 0--1$\deg$ another oblong
(1.6\,$\times$\,0.3$\deg$) emission nebula, Sh\,2-205, at a distance of
900 pc (Fich \& Blitz 1984) is located.  Its ionizing star is HD 24431
(O9\,III).  At its left edge a relatively small nebula Sh\,2-206 is
seen, this is an object of the Perseus arm.  On deep exposures in an
H$\alpha$ filter an emission nebula of about 1.5\,$\times$\,2$\deg$ is
seen at $\ell$, $b$ = 152$\deg$, +4$\deg$, around the star DL Cam (a
triple system, the brightest component is of spectral type B0\,III).

These nebulae and the surrounding areas exhibit evidence of recent star
formation.  The most evident confirmation of this process is the
presence here of a group of numerous young and massive stars known as
the Cam OB1 association located at a distance of $\sim$\,1 kpc.  Its
members are scattered between 132$\deg$ and 156$\deg$ in longitude and
between +7.5$\deg$ and --4$\deg$ in latitude.  It contains more than 40
O--B3 luminosity V--III stars and a few supergiants (Humphreys 1978;
Humphreys \& McElroy 1984; Strai\v{z}ys \& Laugalys 2007a), as well as
the open cluster NGC 1502.  For more about the association see Section 6.

\begin{figure}[tb]
\centering
\includegraphics[width=0.9\textwidth,draft=False]{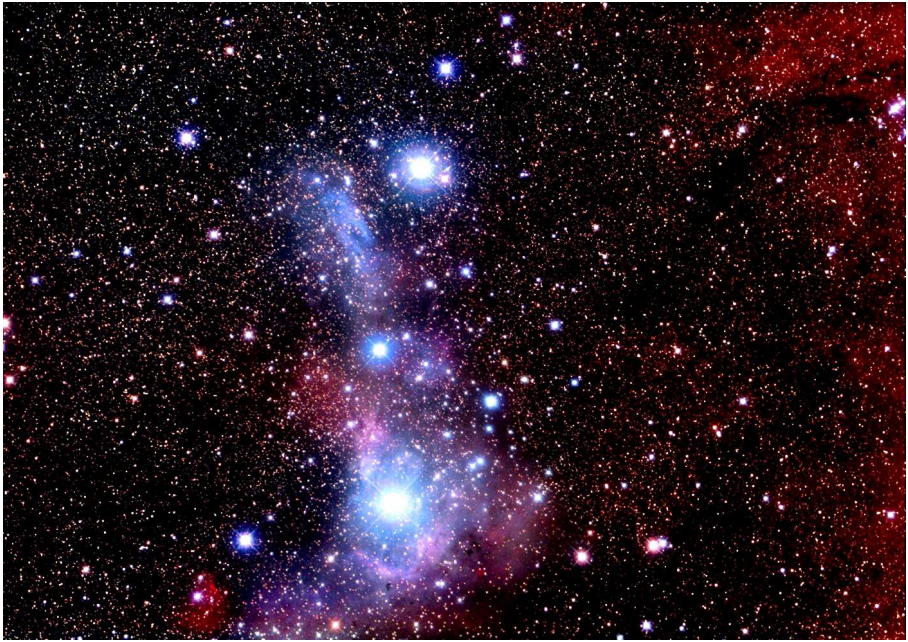}
\vspace{0mm}
\caption{The reflection nebulae vdB\,14 and vdB\,15 at the center of the
Cam OB1-A association.  North is up, east is left.  The
figure is $\sim$\,2$\deg$ high.  Courtesy Adam Block and Tim Puckett.}
\end{figure}

The Cam OB1 association on its western side partly overlaps the
associations Cas OB6 and Per OB1 with several well known objects -- the
H\,II regions W3/4/5 and the double cluster h+$\chi$ Per.  The areas of
these two associations overlap with Cam OB1 only in projection:  in
space they are much farther, in the Perseus arm.  One more group of O
and early B stars, seen in the far background (in or near the Outer
arm), is the Cam OB3 association at $\ell$, $b$ = 146.3--147.7$\deg$,
+2.0--+3.9$\deg$.  The Per OB3 association (with the brightest star
$\alpha$ Per), located in the foreground of Cam OB1, is seen in the
lower part of Figure 1.

Thirteen open clusters, belonging to the Local arm, are present in the
area, their list is given in Strai\v{z}ys \& Laugalys (2007a).  In the
context of this review the most important cluster is NGC 1502 located at
$\ell$, $b$ = 143.7$\deg$, +7.7$\deg$, in the upper part of Figure 1. In
the first lists of associations, this cluster was considered as a
separate association Cam II, since it has two stars of spectral classes
B0\,III (sometimes classified as O9).  Blaauw (1961) suspects that the
run-away star $\alpha$ Cam (O9.5\,Ia) is also related:  it has escaped
from NGC 1502 about 2.0 million years ago.  Later on, stars of this
cluster were included into the list of the Cam OB1 members (Ruprecht
1964; Humphreys \& McElroy 1984).  The ages of both cluster and
association are similar ($\leq$\,10$^7$ yr) since the spectral types of
the most massive stars are $\sim$\,O9 in both of them.

Another cluster, related to the Cam OB1 association, is NGC 1444 at
$\ell$, $b$ = 148.1$\deg$, --1.3$\deg$, in the vicinity of the Sh\,2-205
nebula.  Its appearance is dominated by two stars -- HD 23675 and HD
23800 of 6.7 and 6.9 visual magnitudes and B0.5\,III and B1\,IV spectral
types.  Both of them are in the list of the Cam OB1 members.  Although
Pe\~{n}a \& Peniche (1994) find that the distances of the cluster and
the association are close to each other, there are some doubts whether
NGC 1444 is a real cluster.

A compact group of stars at the southern edge of the H\,II region W4 is
known as the open cluster Stock 7. Its distance and age (700 pc and
16\,$\times$\,10$^6$ yr) given in the Webda database \footnote
{http://www.univie.ac.at/webda} are not very different from those of the
Cam OB1 association.  More details are given in Moffat \& Vogt (1973).

Other clusters in the area having similar distances as the association
probably are not related to it since their ages are between
(150--630)\,$\times$\,10$^6$ yr.  Several clusters of infrared objects
in the area were identified by Carpenter et al.  (2000), Bica et al.
(2003a,b) and Froebrich et al.  (2007a): all of them belong to the
Perseus arm.

\section{Dark Clouds} 

During the last decade several detailed maps of dust distribution along
the Galactic plane were published.  The investigation of Schlegel et al.
(1998) is based on the thermal dust emission survey at 100 $\mu$m by the
IRAS and COBE satellites.  The Dobashi et al.  (2005) atlas is based on
star counts in the Palomar DSS I database charts.  The Froebrich et al.
(2007b) maps are based on the average infrared color excesses in the
2MASS survey.  In Strai\v{z}ys \& Laugalys (2007a), we compared dust
distributions given in these three surveys and found very similar dust
cloud patterns in the area.  The distribution of molecular clouds taken
from the whole sky CO survey by Dame et al.  (2001) is also in a good
agreement.  For a comparison of distributions of dust clouds and young
stars we decided to use the Tokyo atlas (Dobashi et al. 2005), which
better represents dust distribution in the Local arm.  Cloud numbers of
this atlas are designated by TGU, their identification chart is
given in Strai\v{z}ys \& Laugalys (2007a).

In Figure 1 dark clouds are shown as red areas of different density.
The largest extinction is detected between the longitudes 142--143$\deg$
at the latitudes 0--2$\deg$, i.e., at the southern edge of the complex
of the reflection nebulae vdB\,14 and vdB\,15 and about 2$\deg$
southeast of the emission nebula Sh\,2-202.  Here (at $\ell$, $b$ =
142.1$\deg$, +1.6$\deg$) is the clump P1 of the cloud TGU~942, in which
a young massive stellar object (YSO), GL\,490, is immersed (more about
it see in Section 6).  The cloud TGU~942 in the Tokyo atlas is split
into 20 clumps numerated from P1 to P20 which extend from 142$\deg$ to
148$\deg$ at the Sh\,2-205 nebula.  While the densest clumps at the
object GL\,490 are at the same distance as the Cam OB1 association, the
other clumps in TGU~942 belong to different cloud layers.

The dust clouds near the Galactic equator, located within $b$ = 0$\deg$
and +2$\deg$ right of the Sh\,2-202 nebula, belong to the Perseus arm at
a distance of 2.0--2.2 kpc from the Sun.  These clouds are related to
the H\,II regions W3, W4 and W5 and the association Cas OB6.  In the
Tokyo atlas all these clouds are shown under one number TGU~879.

A dense cloud TGU~878 (LDN 1355/58) is seen in the right upper corner of
Figure 1 at $\ell$, $b$ = 133$\deg$, +9$\deg$.  Its lower part belongs
to the Gould Belt layer while the upper part to the Cam OB1 layer.  The
Gould Belt clouds at a distance of 260 pc are illuminated by several
B8--A3 stars and the cepheid SU Cas (Turner \& Evans 1984).

At the longitudes $>$\,148$\deg$ dust clouds form the three possible
configuration versions of ring-like structures described in Strai\v{z}ys
\& Laugalys (2007a).  All of these versions include a curved chain of
small dark clouds TGU 994.  In the Lynds (1962) catalog this chain
includes the clouds from LDN\,1390 to LDN\,1406.  Most noticeable is the
almost perfect ring (or bubble) with the center at $\ell$, $b$ =
152$\deg$, +0.5$\deg$ (at the open cluster NGC 1528) and a diameter of
$\sim$\,8$\deg$.  This ring includes the following Tokyo clouds
(counterclockwise):  TGU~942 (clumps P7 and P8), TGU~994, TGU~1003,
TGU~1036, TGU~1041, TGU~1027 (with $\mu$ Per in foreground), TGU~1014,
TGU~1006 and TGU~989.  According to radial velocities of the associated
CO clouds, all these dust clouds belong to the Local arm, but probably
are located in different layers.  If these clouds form a real bubble,
they should have peculiar motions additional to the Galactic rotation.

\section{Interstellar Extinction from Photometry}  

Until 1996 the interstellar extinction in the Camelopardalis section of
the Milky Way was investigated only scantily.  Probably the first
studies were published by Heeschen (1951), McCuskey (1952) and Kharadze
(1952).  The first of them was based on star counts and two-color
photographic photometry, the next two were based on low-disper\-sion
spectra and two-color photographic photometry.  The conclusion was that
the extinction starts rising at a distance of 100--200 pc.  Rydstr\"om
(1978) investigated the extinction in some Camelopardalis areas by a
spectrophotometric method, and found zero extinction at 100 pc with a
rise up to 2.0 mag at 1 kpc.  However, the above methods were based on
photographic techniques and, consequently, were of low accuracy.

Crude estimates of the extinction in large areas of Camelopardalis near
the Galactic plane have been done by FitzGerald (1968) and Neckel \&
Klare (1980) applying {\it UBV} photometry data and MK spectral types
collected from the literature.  The extinction was found to grow up to
3.5 mag at 1 kpc with no increase at larger distances.  Similar
extinction values were obtained for several open clusters investigated
by two- or three-color photometry.

\begin{figure}[!th]
\centering
\includegraphics[width=0.75\textwidth,draft=False]{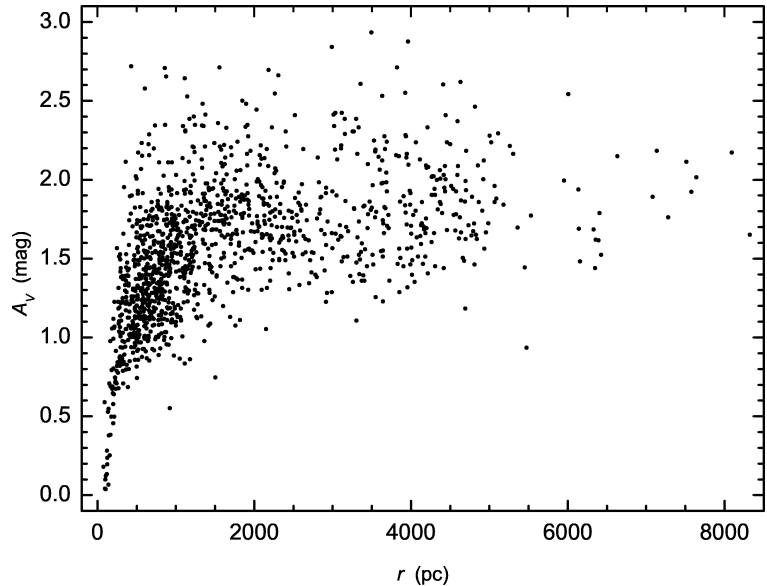}
\vspace{0mm}
\caption{The dependence of interstellar extinction $A_V$ on distance in
the area at $\ell$, $b$ = 146$\deg$, +2.6$\deg$ from Zdanavi\v{c}ius et
al. (2005b).}
\end{figure}

Since 1996, several areas in Camelopardalis were investigated in the
seven-color Vilnius photometric system determining spectral
classes, luminosities and interstellar reddenings of stars with the help
of interstellar reddening-free $Q$-parameters (Zdana\-vi\v{c}i\-us et al.
1996, 2001, 2002a,b, 2005a,b).  Interstellar extinction was studied in
four areas of different sizes using both photoelectric and CCD
photometry of about 2000 stars down to 15.5 mag.  Figure 4 shows the
extinction $A_V$ run with distance in a 1.5 square degree area at
$\ell$, $b$ = 146$\deg$, +2.6$\deg$ investigated by Zdanavi\v{c}ius et
al.  (2005b).  A common property of all areas is the extinction rise
after 120--150 pc reaching about 1.5--2.5 mag at 1 kpc.  At larger
distances the extinction values remain at this level in relatively
transparent areas and reach 2.0--3.0 mag in the direction of dark
clouds.  However, the most reddened stars in these areas were too faint
to be measured, consequently, there is a selection effect present.  The
listed investigations show that the front edge of the Camelopardalis
clouds is almost at the same distance, or even closer, as the Taurus
clouds.

The interstellar extinction law in the direction of Galactic longitudes
135--150$\deg$ was investigated by Zdanavi\v{c}ius et al.  (2002c):  in
the optical range (345--660 nm) it was found to be nearly normal,
typical for the low density interstellar dust.  In the ultraviolet
wavelengths shorter than 330 nm the extinction is found to be slightly
larger than the average.  The ratio $R$ is found to be $\sim$\,2.9,
i.e., a little smaller than normal.

\section{Molecular Clouds}  

The first CO survey of the distribution of molecular gas in the second
quadrant of the Galaxy was published by Cohen et al.  (1980), however,
it ended near the Cassiopeia and Camelopardalis border.  The first
composite CO survey of the entire Milky Way by Dame et al.  (1987)
included all the area discussed in the present review but with a
relatively low resolution.  The next investigation was by Digel et al.
(1996) with a higher resolution, but it included only the longitudes
132--144$\deg$ of those covered by the present review (132--158$\deg$).
In the following discussion we will use the combined results of Digel et
al. and of a much broader CO survey but with lower resolution published
in the second entire sky survey by Dame et al.  (2001).  One more high
resolution CO survey was published by Heyer et al.  (1998) and Brunt et
al.  (2003) but it covers only the longitudes $<$\,141.5$\deg$.

\begin{figure}[tb]
\centering
\includegraphics[width=\textwidth,draft=False]{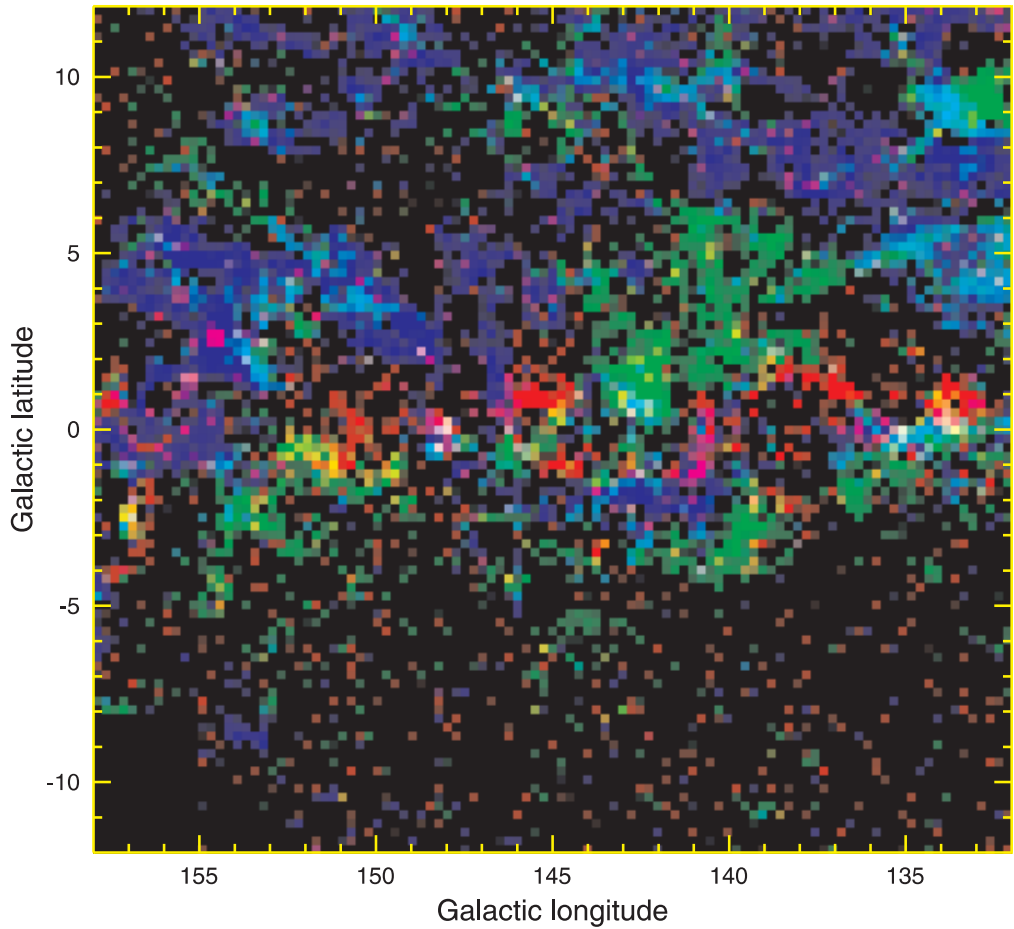}
\caption{CO clouds from Dame et al. (2001). Clouds of the Gould Belt
layer are shown in blue, of the Cam OB1 layer in green and of the
Perseus arm in red. The format of the figure matches that of the
extinction map in Figure~1.}
\end{figure}

Digel et al. conclude that molecular clouds in the investigated
direction are formed by three layers with different radial velocities:
(1) the local layer with velocities between --5 and +10 km/s, (2) the
Cam OB1 layer with velocities between --5 and --20 km/s, and (3) the
Perseus arm with velocities between --30 and --60 km/s.  The space
between the Cam OB1 and the Perseus arm layers is nearly empty in all
the investigated longitude range.  The separation of the local
(hereafter the Gould Belt layer) and the
Cam OB1 layers is not so distinct.  Radial velocities may be transformed
to distances of the layers from the Sun using the Galactic rotation
curve.  Digel et al. accept the mean distance of the local layer
$\sim$\,200 pc and of the Cam OB1 layer $\sim$\,800 pc.  These distances
are close to those determined from stellar photometry.  For the Gould
Belt layer it is more realistic to accept a distance range of 150--300
pc, in accordance with the optical cloud distances in Camelopardalis and
nearby regions (Zdanavi\v{c}ius et al. 2005b; Strai\v{z}ys et al. 2001).
For the second layer the range of distances 900--1000 pc would be in a
better agreement with the optical distance of the Cam OB1 association
(see the next Section).  Both layers contain clouds of different sizes
and densities covering a wide range of longitudes and latitudes.  Many
clouds extend up to +12$\deg$ and some are even at +24$\deg$ from the
Galactic equator (Heithausen et al. 1993; Dame et al. 2001).

In order to assign clouds at larger longitudes to their proper layer,
radial velocities from the Dame et al.  (2001) data have been analyzed
(Strai\v{z}ys \& Laugalys 2007a).  The resulting picture is shown in
Figure 5. The clouds of the Gould Belt layer (150--300 pc) are shown in
blue, of the Cam OB1 layer (900--1000 pc) in green and in the Perseus
arm ($>$\,2 kpc) in red.  It is evident that most clouds seen in the
area belong to the two layers of the Local arm.  The Perseus arm clouds
cover relatively small areas which concentrate near the Galactic
equator.

\vspace{-1mm}

\section{The Cam OB1 Association}  

The association was first identified by Morgan et al.  (1953) and
appeared in the {\it Catalogue of Star Clusters and Associations} (Alter
et al. 1958, 1970).  The size of this association seems to be unusually
large:  its possible members are scattered within 24$\deg$ in longitude
and 12$\deg$ in latitude.  In Strai\v{z}ys \& Laugalys (2007a) we have
suggested that this association may consist of three unrelated groups:
Cam OB1-A around the nebulae vdB\,14, vdB\,15 and Sh\,2-202, Cam OB1-B
around the nebula Sh\,2-205, and Cam OB1-C, the NGC 1502 cluster.  At a
distance of 1 kpc the groups Cam OB1-A and Cam OB1-B would be of about
70--90 pc diameter which is a typical size for OB associations.

Humphreys \& McElroy (1984) list 55 massive member stars of the Cam OB1
association (including two stars of the cluster NGC 1502).  We have
revised this list by estimating photometric distances of the stars from
their MK spectral types and {\em B,V} photometry.  The main source of
distance errors are the accepted luminosity classes.  On the other hand,
some of the potential association stars possess spectral peculiarities
(emission, duplicity, variability, etc.).  Our revised list contains 43
stars plotted in Figure 1:  two O8.5--O9 stars, 35 B0--B3 V--I stars and
six A, G and K supergiants.  About 15 O--B5 stars of the cluster NGC
1502 should be added.  Almost a half of the loose association members
are concentrated in Group A, and only 10 in Group B.

In the area of Group A we also find the Class~I YSO GL\,490, mentioned
in Section 3 and shown in Figure 1. It is known to be a massive object
(8--10 $M_{\odot}$) in a transition stage to Herbig Be stars. surrounded
by a rotating disk and huge envelope, with a $\sim$\,100 AU cavity
inside (see Mitchell et al. 1995; Schreier et al. 2002, 2006).  The dust
envelope gives an extinction $A_V$ of 35\,$\pm$\,5 mag
(Alonso-Costa \& Kwan 1989).

The distance of the Cam OB1 association was estimated by taking the
average of the distances of its members.  For this purpose we chose
stars with reliable MK classification, without peculiarities, and with
{\em BV} photometry available.  The calibration of MK spectral types in
absolute magnitudes $M_V$ was taken from the Strai\v{z}ys (1992)
monograph.  The ratio $R=A_V/E_{B-V} = 2.90$ was used (Zdanavi\v{c}ius
et al. 2002c).  For 26 selected stars the average distance is
1010\,$\pm$\,210 pc.  This value is in good agreement with the earlier
determinations by Humphreys (1978, 1.0 kpc), Melnik \& Efremov (1995,
0.98 kpc), Zeeuw et al.  (1999, 0.9 kpc) and Lyder (2001, 0.98 kpc).

Lyder (2001) has attempted to identify Cam OB1 stars of lower masses
(spectral classes B5--A0 V) in the area of our Group A using spectral,
photometric and radial velocity data from the literature.  He concluded
that star formation in the association has been ongoing during 100
million years.  However, the membership and the evolutionary stage of
the selected stars should be verified by more accurate observations.

\vspace{-1mm}

\section{Low-mass Stars and Infrared Objects}   

The data on young low-mass stars in the Camelopardalis area are quite
scarce.  The Herbig \& Bell (1988) catalog of emission-line stars of
the Orion population lists only one H$\alpha$ emission star, IRAS
03134+5958 ($V$\,$\sim$\,14), located near the questionable open
cluster Stock 23, within the Sh\,2-202 nebula.  The same star, also
known as HBC~336 or CPM\,7, is in the list of YSO objects by Campbell
et al.  (1989).

More young low-mass stars in Camelopardalis have been suspected by Gahm
(1990).  In low-dispersion objective-prism spectra he identified 12
stars with emission in H$\alpha$ in the vicinity of the Sh\,2-205
nebula.  In Strai\v{z}ys \& Laugalys (2007a) we have shown that four of
the Gahm stars in the $J$--$H$ vs.  $H$--$K_s$ diagram are located near
and above the intrinsic line of classical T Tauri (CTTS) and weak-line T
Tauri (WTTS) stars.  The position of the mentioned H$\alpha$ emission
star IRAS 03134+5958 also confirms its relation to T Tauri-type objects.
Two more H$\alpha$ emission stars satisfying the same criterion were
found in the Kohoutek \& Wehmeyer (1997) catalog.  Eleven irregular
variables of types IN and IS, selected from GCVS (Samus et al. 2004) and
listed in Strai\v{z}ys \& Laugalys (2007a), also lie in the same region
of the diagram (Figure~6).

Many YSOs were identified in the Camelopardalis dark clouds using the
available IRAS photometry.  Clemens \& Barvainis (1988) list seven
globules in this area.  Two of them, CB\,17 (LDN 1389) and CB\,26 (LDN
1439) are associated with the IRAS 04005+5647 and IRAS 04559+5200
sources and protostellar cores (Launhardt \& Henning 1997; Launhardt \&
Sargent 2001; Stecklum et al. 2004).  Benson \& Myers (1989) identify 12
dense cores in the LDN 1400 cloud.  Clark (1991) finds in them 32 IRAS
objects.  One of the Clark objects (IRAS 03233+5833) is close to
GL\,490.  Stecklum et al.  (2007) found that the source IRAS 04376+5413,
partially embedded in the small cloud LDN~1415 and associated with the
object HH 892, has brightened considerably in recent years, possibly
indicating that it is an EXor or FUor.  A large nebula surrounding the
binary object IRAS 04261+6339 with a jetlike plume was discovered by
McCall et al.  (2004) at $\ell$, $b$ = 144.5$\deg$, +10.5$\deg$ (see the
object MB\,4 in Figure 1).  At the estimated distance of 2.0\,$\pm$\,0.8
kpc, the object may belong to the Perseus arm.  However, in this case
its distance from the plane (about 370 pc) is unusually large for YSO.
The object in the sky is quite close to the TGU\,951 dust cloud which,
according to the radial velocity of the associated CO cloud, is an
object of the Local arm.

We have increased the number of suspected young stars in the area using
infrared photometry from the 2MASS, IRAS and MSX surveys (Strai\v{z}ys \&
Laugalys 2007b, 2008).  In the $J$--$H$ vs.  $H$--$K_s$ diagram we
isolated infrared objects having $H$--$K_s$\,$\geq$\,0.5 and lying below
the reddening line of O-type stars, which might be reddened YSOs of
Classes I, II and III, see Lada (1987).  From this sample the known
M-type giants of the latest subclasses (including oxygen-rich
long-period variables), OH/IR stars, carbon-rich stars of spectral type
N, Be stars, galaxies and quasars were excluded.  More stars, not
related to star formation, were recognized and excluded according to
their color indices based on IRAS and MSX photometry.  The remaining 187
objects may be considered as potential YSOs.  Evolutionary status was
confirmed for 14 of the brightest objects from this list by obtaining their far
red spectra (Corbally \& Strai\v{z}ys 2008). More suspected YSOs were
confirmed to be H$\alpha$ emission stars by the IPHAS survey (Witham et
al. 2008).


\begin{figure}[!h]
\parbox{70mm}{
\includegraphics[draft=False]{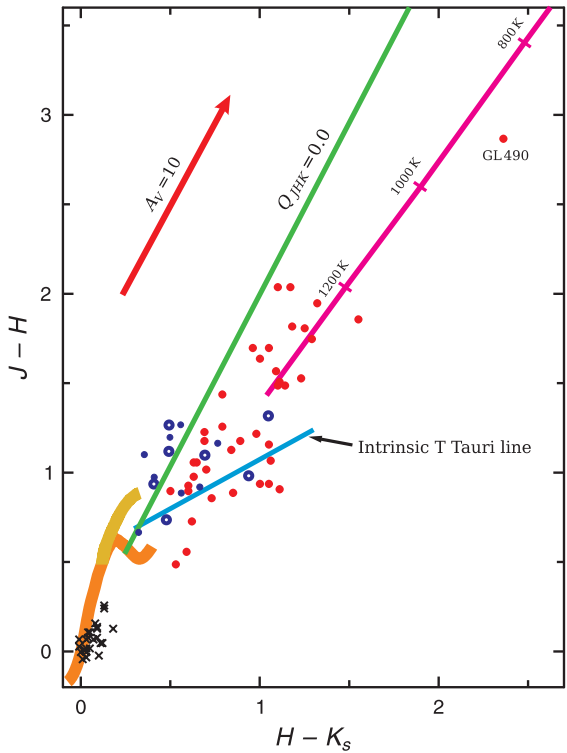}
}
\hspace{.2mm}
\parbox[t]{65mm}{
\vspace{-50mm}
\caption{The $J$--$H$\,vs.\,$H$--$K_s$ diagram for young
stars of the Cam OB1 star-forming region. Crosses designate O--B3
stars of the Cam OB1 association, blue dots designate known irregular
variables, blue open circles are H$\alpha$ emission stars and red dots
are suspected YSOs.  The intrinsic main-sequence and giant lines are
shown in orange and yellow.  The blue line designates the intrinsic
locus of T Tauri stars (Meyer et al. 1997), the violet line is the locus
of black bodies.  The length of the reddening vector (in red)
corresponds to the extinction in the $V$ passband of 10 mag. Infrared
YSOs were searched for in the region below the green line.}
}
\end{figure}
\vskip1mm

The distribution of YSOs in the area exhibits evident clustering in the
darkest dust clouds.  Some groups of objects coincide with known
star-forming regions (W3, W4, W5, LDN 1355/1358 at SU Cas), with the two
centers of the Cam OB1 association (GL\,490 and Sh\,2-205), and with the
infrared clusters described by Bica et al.  (2003a,b) and Froebrich et
al.  (2007a).  The grouping of objects within a few arcminutes around
the object GL\,490 in $K$-band images was first described by Hodapp
(1990, 1994).  However, clustering of infrared objects in the direction
of dense dust clouds does not mean that we are observing a real cluster:
distant heavily reddened K--M giants may imitate it as well
(Strai\v{z}ys \& Laugalys 2008).

The suspected YSOs were attributed to the Gould Belt, the Cam OB1 and
the Perseus arm layers on the basis of radial velocities of the
associated CO clouds. 42 objects were found belonging to the Cam OB1
star-forming region in the Local arm.  In Figure 6 these objects are
plotted in a $J$--$H$ vs.  $H$--$K_s$ diagram together with other young
objects of the same SFR:  O--B3 stars of Cam OB1, irregular variables
and H$\alpha$ emission stars.  In reality, more young objects are
expected above the intrinsic T Tauri line but they are difficult to
identify among thousands of other stars with only 2MASS photometry
available.  Photometric data in the mid- and far-infrared would be
helpful.

The young objects for which fluxes have been measured in the 8--100
$\mu$m wavelength range, can be classified in the Lada (1987) classes
or the Robitaille et al.  (2006, 2007) evolutionary stages.  We
classified about 40 objects observed by IRAS and/or by MSX.  The
spectral energy distribution curves (SEDs) in the $\log\,\lambda
F_{\lambda}$ vs.  $\log \lambda$ coordinates were used (Strai\v{z}ys
\& Laugalys 2007b, 2008).  The object GL\,490 is the brightest among
all, and its SED is typical for a Class I object with a maximum at
50--60 $\mu$m.

\vspace{-3mm}

\section{Conclusions}

The Milky Way in Camelopardalis (including the nearby areas of
Cassiopeia and Per\-seus) was until a decade ago among the least investigated
regions.  However, recently a number of important studies of stars and
interstellar matter in this region have appeared.  This paper is an
attempt to put together the results related to star-forming processes in
the Local spiral arm at the Galactic longitudes from 132$\deg$ to
158$\deg$.

It is evident that this arm is populated by dust and molecular clouds of
high density in which star formation takes place.  The clouds may
be divided into the Gould Belt layer at 150--300 pc distances from the
Sun and the Cam OB1 association layer at $\sim$\,900 pc distance.
However, this division is somewhat ambiguous since in some directions
dust distribution across the arm is almost continuous.  The largest
density is observed in the Cam OB1 layer where some cloud clumps (like
TGU~942~P1) reach an extinction $A_V$ of up to 25 mag or more.

The presence in the area of tens of young massive stars of spectral
classes O--B3 and of supergiants is evidence of a recent star-forming
process which lasted about 10 million years.  Probably this process
continues up to now, since we have succeeded to find in the clouds tens
of stars, which photometrically are similar to YSOs of Classes I--III.
The most typical YSO of Class I is GL\,490, a massive object in a
protostellar evolutionary stage.  The presence of young stars of lower
masses in the clouds is also confirmed by identifying about 17 irregular
variable and H$\alpha$ emission stars.  Most probably, new spectral and
photometric observations will reveal more T Tauri-type stars and related
objects in the clouds.  In the densest parts of the clouds more YSOs in
early evolutionary stages are expected.

The Cam OB1 association sometimes is considered as unusual due to its
enormous linear extent.  However, the association may be split into
three parts -- Group A at the Sh\,2-202 and vdB\,14+15 nebulae, Group
B at the Sh\,2-205 nebula, and a third group formed by the cluster NGC
1502.  Then both the A and B parts will become of normal sizes, 70--90 pc
in diameter.  Both groups contain O--B3 stars, supergiants, H$\alpha$
emission stars and suspected young infrared objects.

\vspace{3mm}

{\bf Acknowledgements}.  We are thankful to T. M. Dame for the
unpublished CO radial velocity data, to G. Gahm for a list of stars with
H$\alpha$ emission, to B. Reipurth for permanent attention and advices
and to Dean Salman, Adam Block and Tim Puckett for use of their pictures
of nebulae.  The use of the 2MASS, IRAS, MSX, Gator, SkyView, Simbad and
Webda databases is acknowledged.

\end{document}